\documentclass[12pt]{iopart}
\usepackage{iopams}
\usepackage{graphicx}
\include{jpeg2ps}
\begin{document}
\title{Warped Angle-deficit of a 5 Dimensional Cosmic String.}{{\it Case I: The General Non-Abelian Case}}
\author{R J Slagter and D Masselink}
\address{Institute of Physics, University of Amsterdam \\ and \\ ASFYON, Astronomisch Fysisch Onderzoek Nederland, Bussum,
The Netherlands }
\ead{info@asfyon.nl}
\begin{abstract}
We present a cosmic string on a warped five dimensional space time  in Einstein-Yang-Mills theory.
Four-dimensional cosmic strings show some serious problems concerning the mechanism of string smoothing related to the string mass per unit length, $G\mu \approx 10^{-6}$. A warped cosmic string could overcome this problem and also the superstring requirement that $G\mu$ must be of order 1, which is far above observational bounds. Also the absence of observational evidence of axially symmetric lensing effect caused by cosmic strings could be explained by the warped cosmic string model we present: the angle deficit of the string is warped down to unobservable value in the brane, compared to its value in the bulk. It turns out that only for negative cosmological constant, a consistent numerical solution of the model is possible.

\section{Introduction}

Recently, there is growing interest in the Randall-Sundrum(RS) warped 5D geometry\cite{Ran1,Ran2}. One of the interesting outcomes of this idea is the solution of the large hierarchy problem between the weak scale and the fundamental scale of gravity. The predicted Kaluza-Klein particles in the model could be detected with the LHC at CERN. In the original RS scenario, it was proposed that our universe is five dimensional, described by the metric
\begin{equation}
ds^2=e^{-2\mid y \mid k y_c}g_{\mu\nu}dx^\mu dx^\nu +y_cdy^2.
\end{equation}
The extra dimension y makes a finite contribution to the 5D volume because of the exponential warp factor, where $y_c$ is the size of the
extra dimension.
At low energies, gravity is localized at the brane and general relativity is recovered. At high energy gravity "leaks"
into the bulk. The 4D Planck scale will be an effective scale which can become much larger than the fundamental Planck scale $M_P$ if
the extra dimension is much larger than $M_P^{-1}$.
Further, the self-gravity of the brane must be incorporated. This will protect the 3 dimensional space from the large extra dimensions
by curvature rather than straightforward compactification.
Also matter fields in the bulk can be incorporated. This will lead to a kind of "holographic" principle, i.e., the 5D dynamics may be determined from knowledge of the fields on the 4D boundary. For an overview, see \cite{Roy}.
We will consider here the 5D model with a general Yang-Mills field, dependent of $ r, y$ and $t$.
In a following article we investigate the interplay of the 4D and 5D coupled equations with the junction conditions.

\section{The model}
We will consider here the RS2 model with two branes at $y=0$, the weak visible brane and at $y=y_c$, the gravity brane (in the RS1 model one let
$y_c\rightarrow \infty$).
The action of the model under consideration is \cite{Okuy}
\begin{eqnarray}
{\cal S}=\frac{1}{16\pi}\int d^5x\sqrt{-^{(5)}g}\Bigl[\frac{1}{ G_5}(^{(5)}R-\Lambda_5)+\kappa\Bigl(^{(5)}R_{\mu\nu\alpha\beta}
 ^{(5)}R^{\mu\nu\alpha\beta}-4 ^{(5)}R_{\alpha\beta} ^{(5)}R^{\alpha\beta}\cr+^{(5)}R^2\Bigr)-\frac{1}{g^2}Tr{\bf F^2}\Bigr]
+\int d^4x\sqrt{-^{(4)}g}\Bigl[\frac{1}{G_5}\Lambda_4+  S_4 \Bigr]
\end{eqnarray}
with $G_5$ the gravitational constant, $\Lambda_5$ the cosmological constant, $\kappa$ the Gauss-Bonnet
coupling, $g$ the gauge coupling, $\Lambda_4$ the brane tension and $S_4$ the effective 4D Lagrangian, which is given by a generic functional
of the brane metric and matter fields on the brane and will also contain the extrinsic curvature corrections due to the projection of the
5D curvature.
For the moment we will consider here only the 5D equation in a general setting and with a Yang-Mills matter field. The 4D induced equations
together with the junction conditions will be presented in part 2 of a next article.

The coupled set of equations of the EYM-GB system will then become( from now on all the indices run from 0..4)
\begin{eqnarray}
\Lambda_5  g_{\mu\nu}+G_{\mu\nu}-\kappa GB_{\mu\nu}=8\pi G_5 T_{\mu\nu},
\end{eqnarray}
\begin{eqnarray}
{\cal D}_\mu F^{\mu\nu a}=0,
\end{eqnarray}
with the Einstein tensor
\begin{eqnarray}
G_{\mu\nu}= R_{\mu\nu}-\frac{1}{2}g_{\mu\nu} R,
\end{eqnarray}
and Gauss-Bonnet tensor
\begin{eqnarray}
GB_{\mu\nu}=\frac{1}{2}g_{\mu\nu}\Bigl( R_{\gamma\delta\lambda\sigma}R^{\gamma\delta\lambda\sigma}
-4R_{\gamma \delta}R^{\gamma\delta} +R^2 \Bigr) -2RR_{\mu\nu}+4R_{\mu\gamma}{R^{\gamma}}_{\nu} \cr +
4R_{\gamma\delta}{{{R^{\gamma}}_{\mu}}^{\delta}}_{\nu}
 -2R_{\mu\gamma\delta\lambda}{R_{\nu}}^{\gamma\delta\lambda}.
\end{eqnarray}
Further, with
$R_{\mu\nu}$ the Ricci tensor and $T_{\mu\nu}$ the energy-momentum tensor
\begin{eqnarray}
T_{\mu\nu}={\bf Tr}F_{\mu\lambda}F_\nu^\lambda -\frac{1}{2}g_{\mu\nu}{\bf Tr}F_{\alpha\beta}F^{\alpha\beta},
\end{eqnarray}
and with $F_{\mu\nu}^a=\partial_\mu A_\nu^a -\partial_\nu A_\mu^a +g\epsilon^{abc}A_\mu^b A_\nu^c $, and
${\cal D}_\alpha F_{\mu\nu}^a=\nabla_\alpha F_{\mu\nu}^a+g\epsilon^{abc}A_\alpha^b F_{\mu\nu}^c$
where $A_\mu^a$ represents  the YM potential.

We will consider the warped axially symmetric space time
\begin{equation}
ds^2=-F(t,r,y)[dt^2-dz^2-dr^2-A(t,r,y)d\varphi^2]+dy^2,
\end{equation}
with y the bulk dimension and the YM parameterization
\begin{eqnarray}
A_t^{(a)}=\Bigl(0,0,\Phi(t,r,y)\Bigr),\quad A_r^{(a)}=A_z^{(a)}=A_y^{(a)}=0, \cr  A_\phi^{(a)}=\Bigl(0,0,W(t,r,y)\Bigr).
\end{eqnarray}
So the metric and YM components depend t and the two space dimensions r and y.

The set of PDE's become, for $\kappa =0$ for the time being,
\begin{equation}
F_{tt}=F_{rr}+\frac{1}{2}FF_{yy}+\frac{3}{4F}(F_t^2-F_r^2)+\frac{1}{2}\Lambda F^2-\frac{4\pi G}{A}\Bigl[W_r^2-W_t^2+FW_y^2\Bigr],
\end{equation}
\begin{eqnarray}
A_{tt}=A_{rr}+FA_{yy}-\frac{1}{2A}(A_r^2+A_y^2-A_t^2)+\frac{1}{F}(F_rA_r-F_tA_t+2FF_yA_y)\cr
-\frac{16\pi G}{F}\Bigl[W_t^2-W_r^2-FW_y^2\Bigr],
\end{eqnarray}
\begin{equation}
W_{tt}=W_{rr}+FW_{yy}+\frac{1}{2A}W_tA_t+W_y(F_y-\frac{F}{2A}A_y)-\frac{1}{2A}W_rA_r.
\end{equation}

\section{ The Static case}

In the static case the resulting PDE's become
\begin{equation}
A_{rr}+FA_{yy}+2F_yA_y+\frac{F_rA_r}{F}-\frac{FA_y^2+A_r^2}{2A}+\frac{16\pi G}{F}\Bigl(W_r^2+FW_y^2\Bigr)=0,
\end{equation}
\begin{equation}
F_{rr}+FF_{yy}+F_y^2+\frac{F_rA_r+FF_yA_y}{2A}+\frac{2}{3}\Lambda F^2-\frac{16\pi G}{3A}\Bigl(W_r^2+FW_y^2\Bigr)=0,
\end{equation}
\begin{equation}
W_{rr}+FW_{yy}-\frac{W_rA_r}{2A}+W_y(F_y-\frac{FA_y}{2A})=0,
\end{equation}
\begin{equation}
\Phi_{rr}+F\Phi_{yy}+\frac{\Phi_rA_r}{2A}+\Phi_y(F_y+\frac{FA_y}{2A})=0.
\end{equation}
We also have the two constraints
\begin{equation}
F\Phi_y^2+\Phi_r^2=0, \quad W_r\Phi_r+FW_y\Phi_y=0.
\end{equation}
When we substitute the equations for $\Phi$ and W into the conservation equation $\nabla_\mu T^{\nu\mu}=0$, we obtain identically zero, as it should be.
When we introduce the quantities $\theta_i$ defined by
\begin{eqnarray}
\theta_1\equiv\frac{F}{\sqrt{A}}A_r, \quad  \theta_2\equiv\sqrt{A}F_r, \quad \theta_3\equiv\frac{F^2}{\sqrt{A}}A_y, \quad
\theta_4\equiv F\sqrt{A}F_y,
\end{eqnarray}
then the equations can be written as
\begin{equation}
\frac{\partial}{\partial r}\theta_1+\frac{\partial}{\partial y}\theta_3=-\frac{16\pi G}{\sqrt{A}}(W_r^2+FW_y^2),
\end{equation}
\begin{equation}
\frac{\partial}{\partial r}\theta_2+\frac{\partial}{\partial y}\theta_4=\frac{16\pi G}{3\sqrt{A}}(W_r^2+FW_y^2)-\frac{2}{3}
\Lambda F^2\sqrt{A},
\end{equation}
\begin{equation}
\Bigl[\frac{W_r}{\sqrt{A}}\Bigr]_r+\Bigl[\frac{FW_y}{\sqrt{A}}\Bigr]_y=0,
\end{equation}
\begin{equation}
\Bigl[\sqrt{A}\Phi_r\Bigr]_r+\Bigl[F\sqrt{A}\Phi_y\Bigr]_y=0.
\end{equation}

The Ricci scalar $^{(5)}R$ becomes:
\begin{equation}
^{(5)}R=\frac{8\pi G}{3AF^2}\bigl(FW_y^2+W_r^2\Bigr)+\frac{5}{3}\Lambda_5.
\end{equation}
We now investigate the properties of the static solution for large values of $r$ and $y$. We will assume that
\begin{equation}
\int_{0}^{\infty}\sqrt{A}\sigma dr
\end{equation}
converges, where $\sigma$ is the energy density $T_0^0 = -4\pi G\frac{W_r^2+FW_y^2}{2AF^2}$.
Further,
\begin{equation}
\lim_{r\rightarrow \infty}\sqrt{A}\sigma =0.
\end{equation}

\section{ Analysis of the Angle Deficit}
The angle deficit can be calculated for a class of static translational symmetric space times which are asymptotically Minkowski minus a wedge.
If we denote with $l$ the length of an orbit of $\Bigl(\frac{\partial}{\partial\varphi}\Bigr)^a$ in the brane, then the angle deficit is given
by\cite{Vil,Garf,Ford Vil}
\begin{equation}
(2\pi -\Delta\varphi )=\lim_{r\rightarrow\infty}\frac{d l}{dr},
\end{equation}
with
\begin{equation}
l=\int_0^{2\pi}\sqrt{g_{ab}\Bigl(\frac{\partial}{\partial\varphi}\Bigr)^a\Bigl(\frac{\partial}{\partial\varphi}\Bigr)^b} d\varphi.
\end{equation}

One better can use the Gauss-Bonnet theorem to obtain the angle deficit by calculating the integral of the Gaussian curvature over the surface of
$S(t,z)$ = const. If one transports a vector around a closed curve, then the angle rotation $\alpha$ will be given\cite{Ford Vil} by the area integral
over of the subsurface of S

\begin{equation}
\alpha  =\int d^3 x\sqrt{^{(3)}g}^{(3)}K,
\end{equation}
with
\begin{equation}
^{(3)}K=\frac{1}{2}{^{(3)}g}^{ik} {^{(3)}g}^{jl}{^{(5)}R}_{ijkl}.
\end{equation}

For our case, we obtain
\begin{eqnarray}
\sqrt{^{(3)}g}{^{(3)}K}=-\frac{1}{2}\Bigl(\frac{F_r\sqrt{A}}{F}\Bigr)_r-\frac{1}{2}\Bigl(\frac{A_r}{\sqrt{A}}\Bigr)_r
-\frac{1}{2}\Bigl(\frac{FA_y}{\sqrt{A}}\Bigr)_y-(\sqrt{A}F_y)_y\cr+\frac{1}{4}\sqrt{A}F_y\Bigl(\frac{F_y}{F}+\frac{A_y}{A}\Bigr)\cr
=-\frac{1}{4}\Bigl[\frac{F_y^2}{F^2}+\frac{F_r^2}{F^3}-\frac{2}{3}\Lambda -\frac{80\pi G}{3F^2A}(W_r^2+FW_y^2)\Bigr].
\end{eqnarray}
Then Eq.(28) becomes
\begin{eqnarray}
\alpha=-\pi \Bigl\{\Big[\Bigl(\frac{\sqrt{A}F_r}{F}\Bigr)+\Bigl(\frac{A_r}{\sqrt{A}}\Bigr)\Bigr]^\infty_{r=0}
+\Bigl[\frac{FA_y}{\sqrt{A}}+2\sqrt{A}F_y\Bigr]^\infty_{y=0}\cr
-\frac{1}{2}\int_0^\infty \int_0^\infty \sqrt{A}F_y(\frac{F_y}{F}+\frac{A_y}{A})drdy\Bigr\}.
\end{eqnarray}
Or, using the second expression in Eq.(30)
\begin{eqnarray}
\alpha =-\frac{1}{2}\pi\int_0^\infty\int_0^\infty F\sqrt{A}\Bigl(\frac{20}{3}\sigma -\frac{2}{3}\Lambda\Bigr)drdy \cr -\frac{1}{2}\pi\int_0^\infty\int_0^\infty F\sqrt{A}\Bigl[\frac{F_y^2}{F^2}+\frac{F_r^2}{F^3}\Bigr]drdy.
\end{eqnarray}
If one assumes that in the 4 dimensional case, for $r\rightarrow\infty: F\rightarrow 1$ and $\sqrt{A}\rightarrow br$ and
for $r\rightarrow 0: F\rightarrow 1$ and $\sqrt{A}\rightarrow r$, than the  first term of Eq.(31) represents the well-known result \cite{Garf}
that $\alpha =2\pi(1-b)$ in de brane, so S is asymptotically a conical surface. In the 5 dimensional case the results depend on the boundary values of our warp factor F, i.e., the last two terms in Eq. (31). From Eq.(32) one observes that the first term represents the proper mass per unit length
of the string plus a contribution from the cosmological constant. The second term is the correction term.

Now we try to obtain for the asymptotic warped metric
\begin{equation}
ds^2=F_c e^{k_2 y+a_2}\Bigl[-dt^2+dz^2+dr^2+(k_1r+a_1)^2d\varphi^2\Bigr]+dy^2.
\end{equation}
For $y=0$ we recover de 4D result of a flat space time minus a wedge by the transformation\cite{Garf}
\begin{eqnarray}
r'=r+\frac{a_1}{k_1},\qquad \varphi '=k_1\varphi \quad( 0\leq \varphi \leq 2\pi ),
\end{eqnarray}
i.e.,
\begin{equation}
ds^2=-dt^2+dz^2+d(r')^2+(r')^2d(\varphi ')^2+dy^2,
\end{equation}
where now $\varphi '$  has a different range then $\varphi$.
For $y\neq 0$ we have the warped metric
\begin{equation}
ds^2=F_ce^{k_2 y+a_2}\bigl[-dt^2+dz^2+d(r')^2+(r')^2d(\varphi ')^2\Bigl] +dy^2.
\end{equation}
The angle deficit is determined by $k_1F_ce^{k_2y+a_2}$.

Let us consider now
\begin{eqnarray}
\frac{\partial }{\partial r}(\theta_1+\theta_2)+\frac{\partial}{\partial y}(\theta_3+\theta_4)=
-\frac{32\pi G}{3\sqrt{A}}\Bigl(W_r^2+FW_y^2\Bigr)-\frac{2}{3}\Lambda\sqrt{A}F^2 \cr
 =-\frac{32\pi G}{3}\Bigl[\frac{\partial}{\partial r}\Bigl(\frac{WW_r}{\sqrt{A}}\Bigr)+\frac{\partial}{\partial y}\Bigl(\frac{F W W_y}{\sqrt{A}}\Bigr)\Bigr]
-\frac{2}{3}\Lambda \sqrt{A}F^2,
\end{eqnarray}
where we used the Eq.'s (19)-(22).
After rearranging we then obtain
\begin{equation}
\frac{\partial }{\partial r}\Bigl(\theta_1+\theta_2+\frac{32}{3}
\pi G\frac{WW_r}{\sqrt{A}}\Bigr)=-\frac{\partial }{\partial y}\Bigl(\theta_3+\theta_4+\frac{32}{3}
\pi G\frac{FWW_y}{\sqrt{A}}\Bigr)-\frac{2}{3}\Lambda\sqrt{A}F^2.
\end{equation}
So we notice that the $\Phi$-field disappears from the equation. It will have only a contribution on the brane.
It is quite easy to obtain a particular  solution of this equation, Eq.(38). For
\begin{eqnarray}
F=F_c e^{\pm\sqrt{-\Lambda} y+a_2}\qquad A=A_c(k_1 r+a_1)^2 ,
\end{eqnarray}
we obtain for $W$ a solution of the form $W(r,y)=W_1(r)W_2(y)$, where $W_1$ and $W_2$ are given by Bessel functions.
This oscillatory behavior of $W$ is not uncommon for gravitating YM vortices.

So it seems to be possible to find the desired asymptotic warped form for the conical space time, i.e., Eq.(33).

The next task is to obtain from the junction condition and the brane-bulk splitting, relations between the several
constants in the model.

\section{Numerical solutions}

For a given set of initial conditions, these PDE's determine the behavior of F, A, W and $\Phi$. We will impose particular asymptotic conditions,
in order to obtain acceptable solutions of the cosmic string. First, we have
$\lim_{r\rightarrow\infty} \Phi (r,y)=1, \lim_{r\rightarrow\infty} W (r,y)=0, F(0,y)=1, A(0,y)=0, \frac{\partial}{\partial r}A(0,y)=1 $.
Further, $\lim_{r\rightarrow\infty}\frac{g_{\varphi\varphi}}{r^2}=1$.

We solve the system using the numerical code CADSOL-FIDISOL.
The asymptotic form of the metric component $g_{\varphi\varphi}$ behaves as expected.

\begin{figure}
 \includegraphics[width=5cm,bb=0 0 320 320]{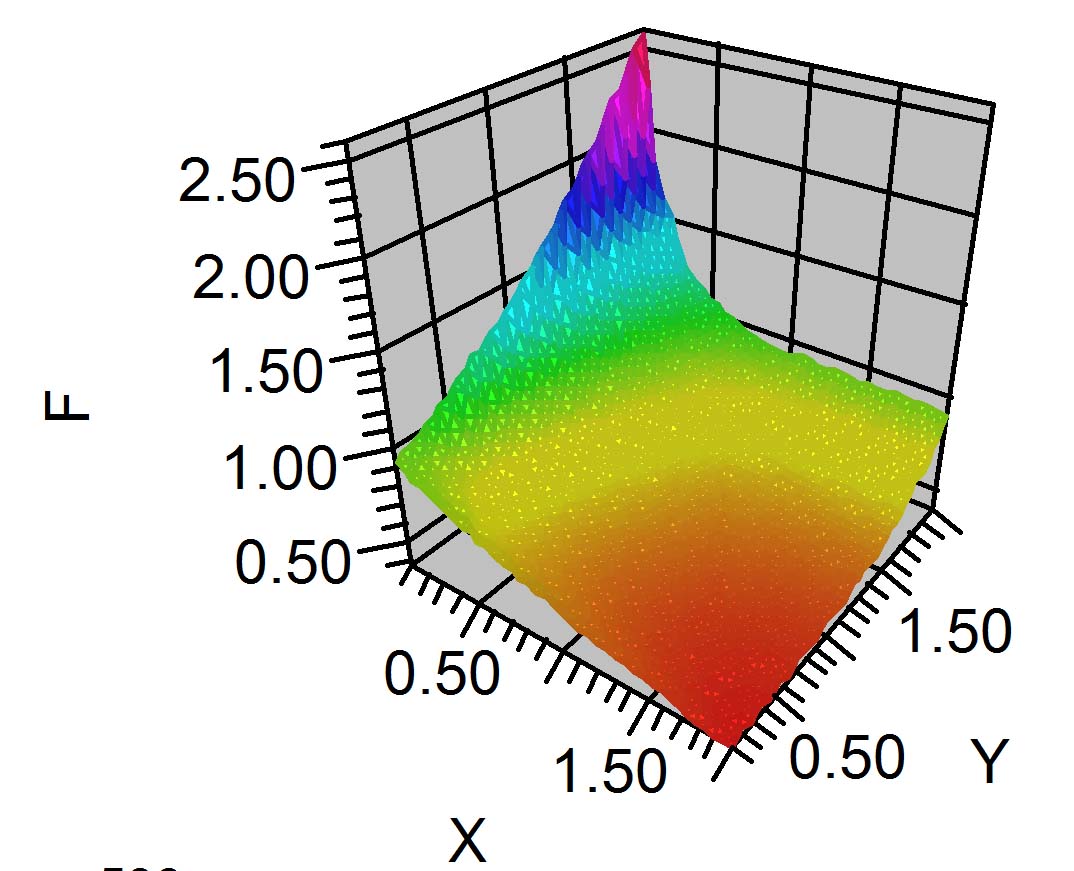}
 \includegraphics[width=5cm, bb=0 0 320 320]{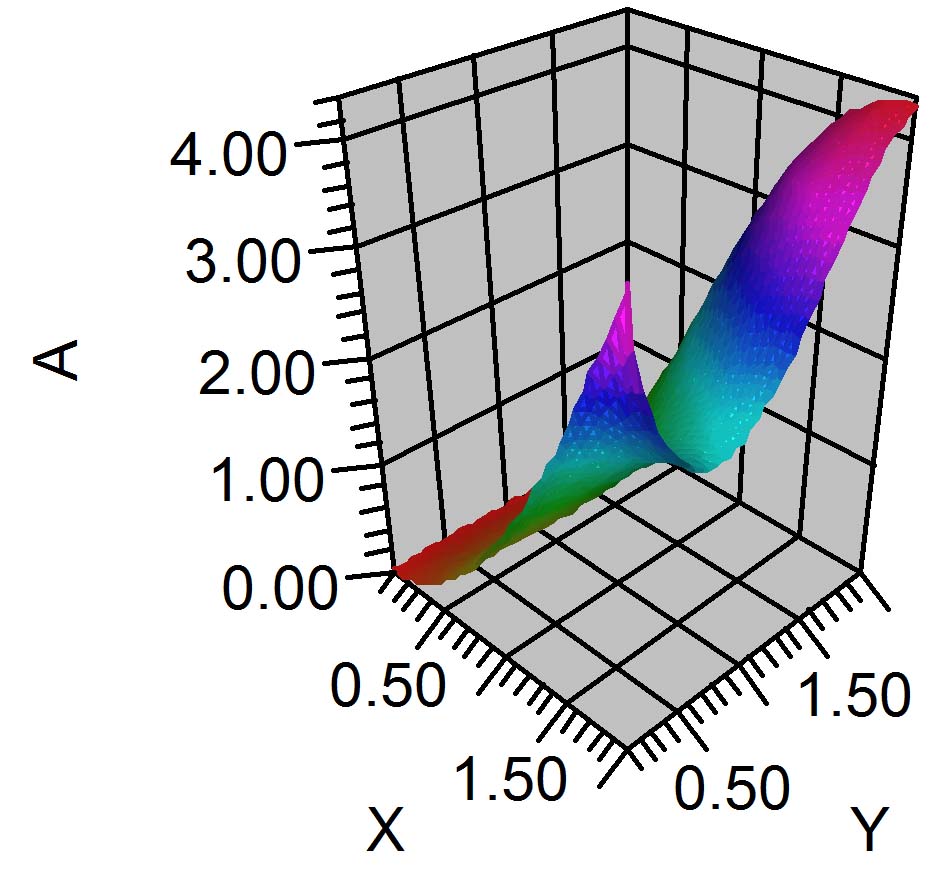}
  \includegraphics[width=5cm, bb=0 0 320 320]{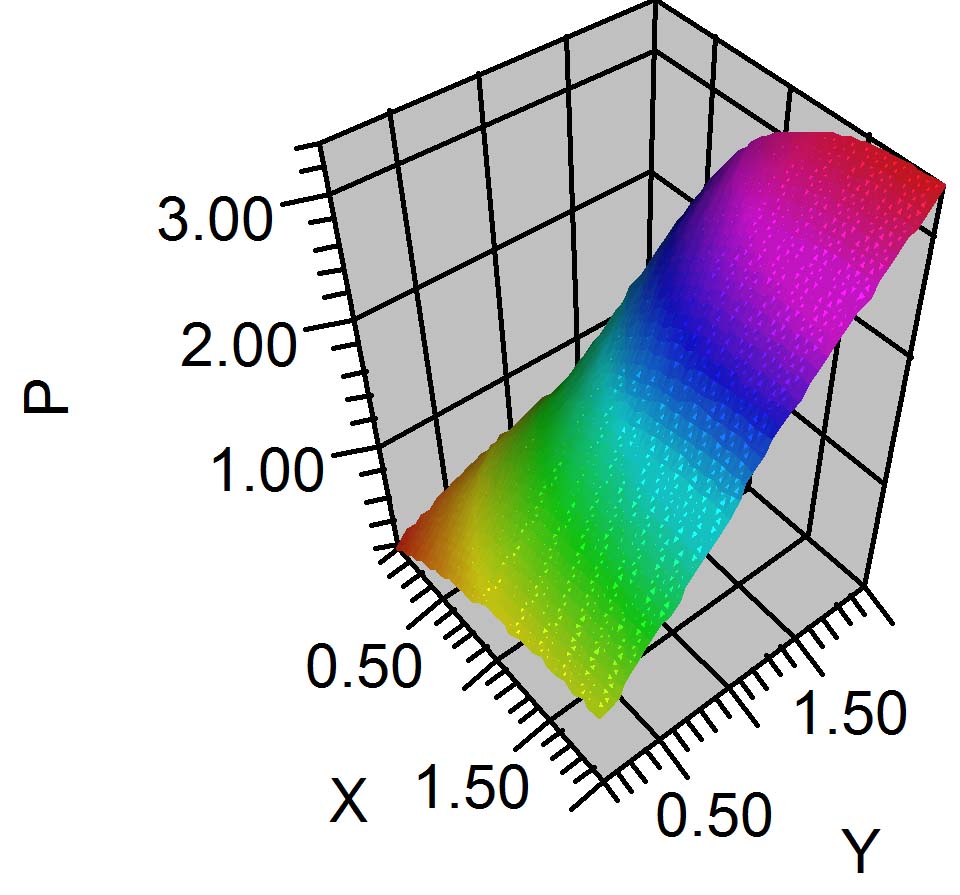}
 \includegraphics[width=5cm, bb=0 0 320 320]{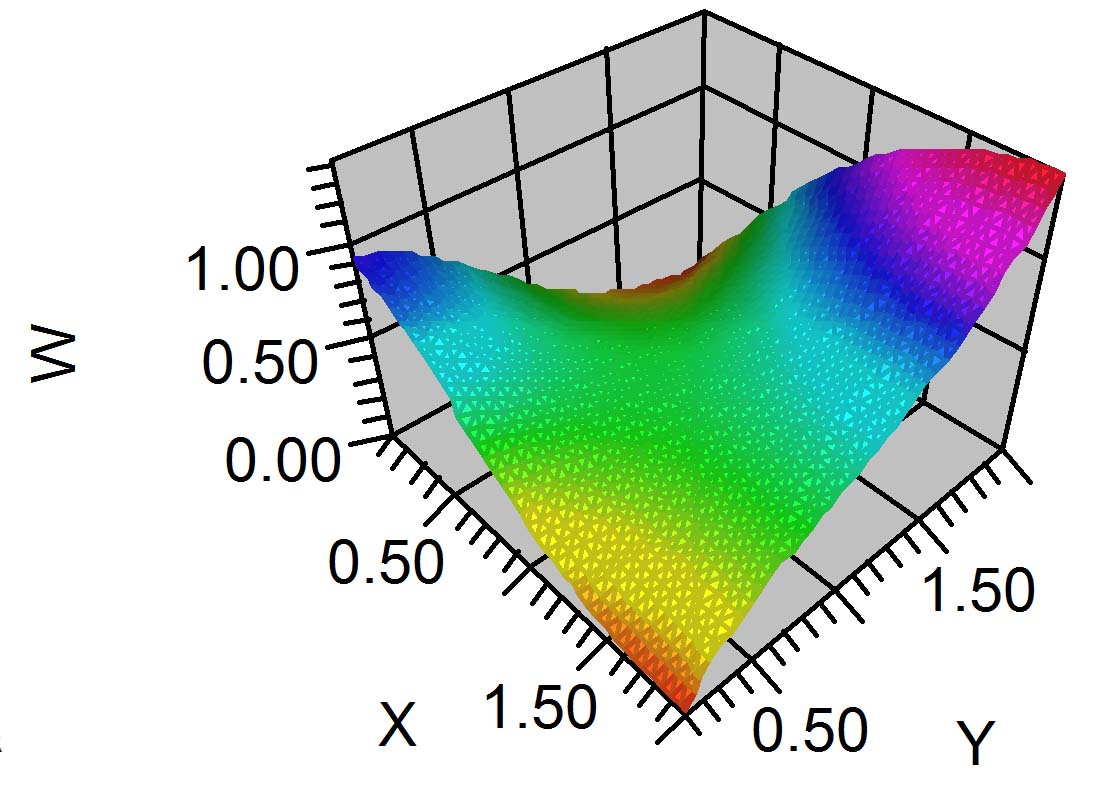}
 \includegraphics[width=5cm, bb=0 0 320 320]{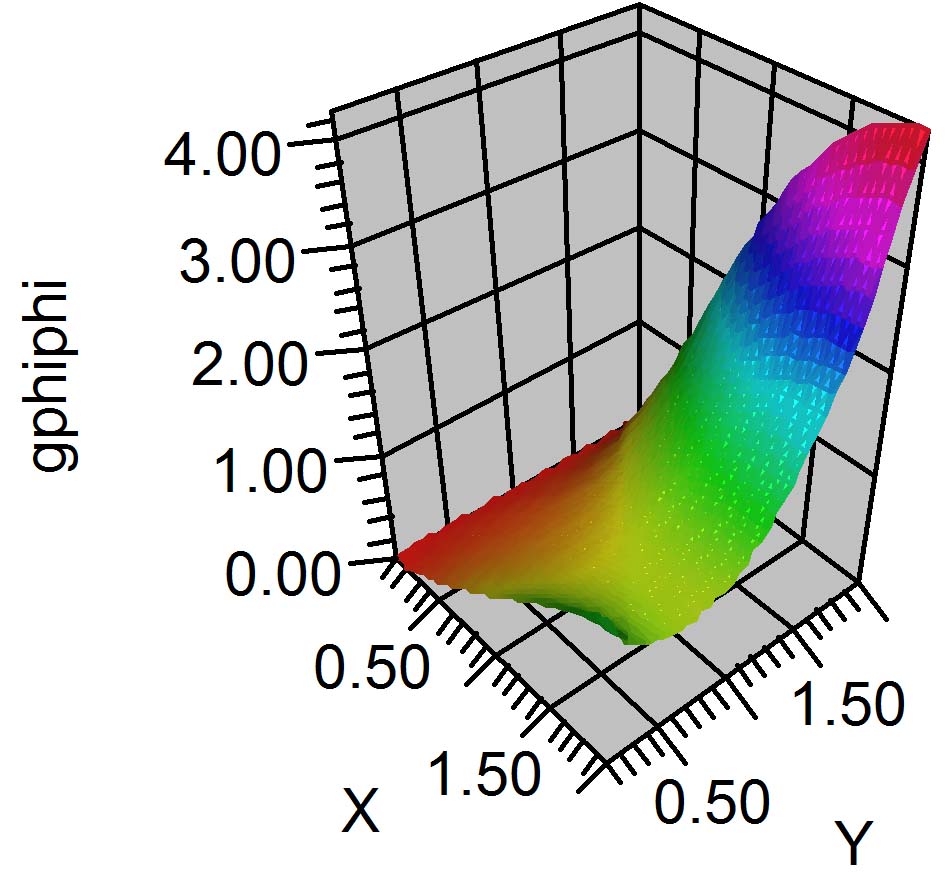}
\caption{Typical solution of F, A, W and $\Phi$ for negative $\Lambda$ with initial values: $ F=e^{(-r+y)}, A=r^2,
W=e^{(-r^2-y^2)}, \Phi = 1-e^{(-r-y)}$ and Dirichlet boundary conditions on the outer boundaries. We also plotted $g_{\varphi\varphi}$}
\end{figure}

\section{Conclusions}
In earlier attempts\cite{Slag1,Slag2,Slag3}, we tried to build a 5-dimensional cosmic string without a warp factor and investigated
the causal structure. Here we considered a different approach.
It seems possible that the absence of cosmic strings in observational data could be explained by our model, where the effective angle-deficit resides in the bulk and not in the brane.
In this part we considered the 5D equations in general form, without the splitting of the energy-momentum tensor in a bulk and brane part.
We find a consistent set of equations in the bulk. The asymptotic behavior of the metric outside the core of the string seems to have
the desired form. This solution must be consistent with the system of equations obtained by the bulk-brane splitting.
It is also interesting to investigate holographic ideas in our model.
 These subjects are under study by the authors and will be published in a followup article.
\begin{figure}
\includegraphics[width=8cm, bb=0 0 600 400]{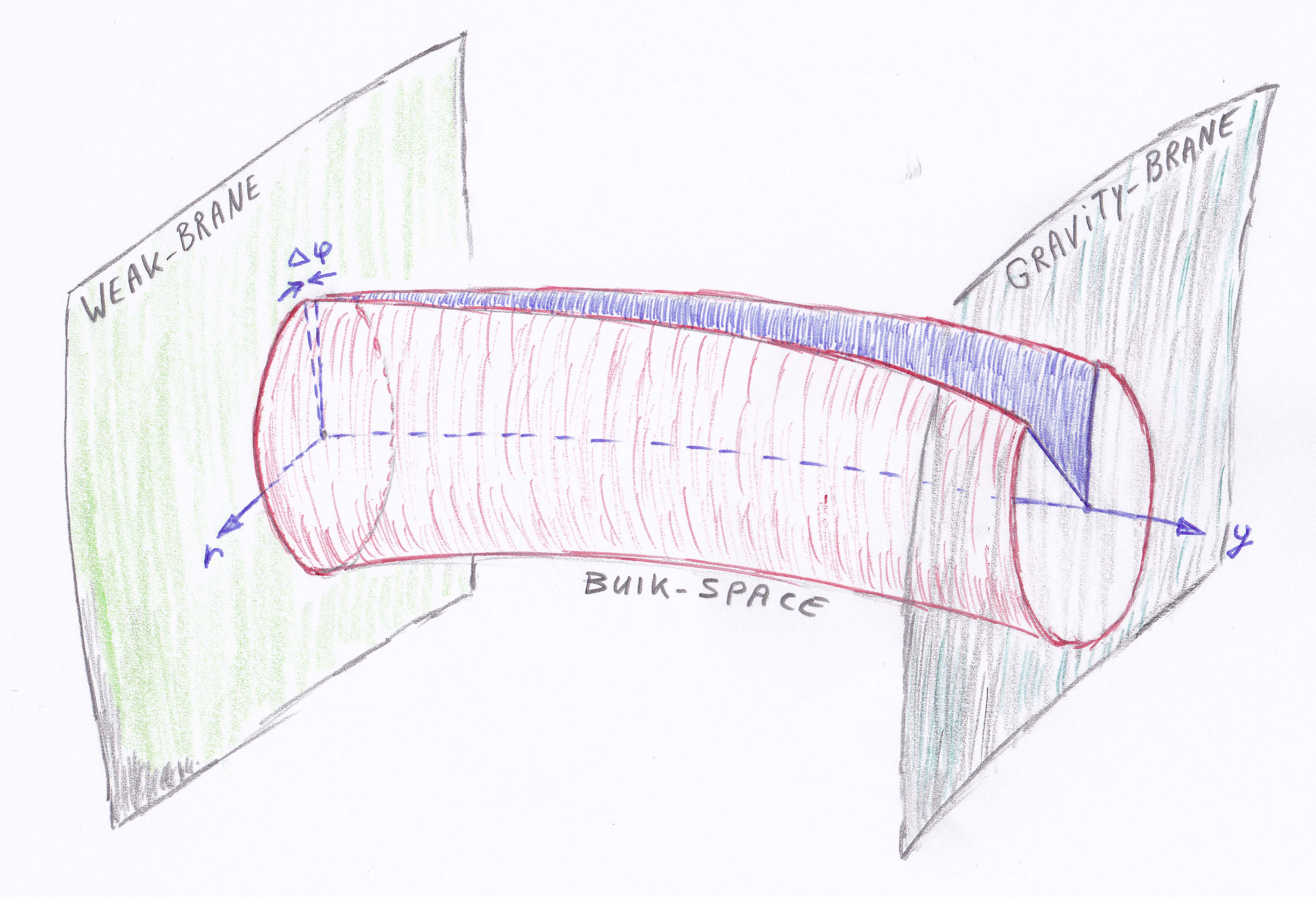}
\caption{The 5-D cosmic string}
\end{figure}

\end{abstract}
\maketitle

\end{document}